\documentstyle[epsfig]{article}
\begin{document}
\begin{large}

\centerline{ N-N Interactions in the Extended Chiral SU(3) Quark
Model
%\end{large}
%\thanks{This work was
%partly supported by the National Nature Science Foundation of China}}
%\par
\footnote{Project supported by the National Natural Science
Foundation of China (No.10047002) and  Scientific Research Foundation
of Liaoning Education Department (No. 202122028)}}
\end{large}
\par
\vspace{0.5cm} \centerline{L. R. Dai}
\begin{small}
\centerline{Department of Physics, Liaoning Normal University,
116029, Dalian, P. R. China}
\end{small}
\vspace{0.2cm} \centerline{Z. Y. Zhang, Y. W. Yu, and P. Wang}
\begin{small}
\centerline{Institute of High Energy Physics, 100039, Beijing,
P. R. China}
\end{small}
%\begin{center}
%\begin{minipage}{120mm}
\vskip 0.5in
%\vskip 1in
%\baselineskip 0.1in
%\def\baselinestretch{5.0}
%\centerline{\bf Abstract}
%\noindent

\begin{abstract}
The chiral $SU(3)$ quark model is extended to include coupling
between vector chiral field and quarks. By using this model,
the phase shifts of NN scattering for different partial
waves are studied.
The results are very similar to those of the chiral SU(3) quark
model calculation, in which one gluon exchange (OGE) plays
dominate role in the short range part of the quark-quark
interactions. Only in the $^1S_0$ case, the one channel phase
shifts of the extended chiral $SU(3)$ quark model are obviously
improved.
\end{abstract}

\vspace{1.0cm}


%\end{titlepage}
%\end{description}
%\newpage

\vspace{1.5cm}
\section{Introduction}

As is well known, in the light quark system the non-perturbative
Quantum Chromodynamics (QCD) effect is important and not
negligible. An effective approach to describe such effect can be
made by introducing the coupling between the chiral fields and
quarks, especially in studying the nucleon-nucleon (N-N)
interactions. A chiral $SU(3)$ quark model \cite{s1,s2} was
proposed by generalizing the idea of the $SU(2)$ $\sigma$ model to
the flavor $SU(3)$ case. In the original chiral $SU(3)$ quark
model, the nonet pseuo-scalar meson exchanges and the nonet scalar
meson exchanges are considered in describing the medium and long
range parts of the interactions, and the one gluon exchange (OGE)
potential is still retained to contribute the short range
repulsion. By using this model, the energies of the baryon ground
states, the N-N scattering phase shifts and the
hyperon-nucleon (Y-N) cross sections can be reproduced
reasonably. It seems that the repulsive core of the N-N
interaction can be explained by the OGE and the quark exchange
effect.

Since last few years, Shen et al \cite{s3}, Riska and Glozman
\cite{s4,s5} applied the quark-chiral field coupling  model to
study the baryon structure. They found that the chiral field
coupling is also important in explaining the structure of baryons.
Especially, the $\pi$  field coupling leads to increase the weight
of D wave components in N and $\Delta$, and improve the mass of
the Roper resonance. In the work of Riska et al \cite{s4,s5}, the
vector meson coupling was also included to replace OGE. They
pointed out the spin-flavor interaction is important in explaining
the energy of the Roper resonance and got a comparatively good fit
to the baryon spectra. Despite the fact that this model can not
explain the decay properties of the baryon excited states in the
framework of three valance quark model space, it is still a big
challenge to the Isgur's model \cite{s6}, in which the OGE governs
the baryon structure.

On the other hand, in the study of baryon-baryon (B-B)
interactions on quark level, the short range repulsive feature of
N-N interaction can be explained by OGE interaction and quark
exchange effect \cite{s7,s8}, but in the traditional one boson
exchange (OBE) model on baryon level \cite{s9} the N-N short range
repulsion comes from vector meson ($\rho,\omega, K^*$ and $\phi$)
exchanges. Some authors \cite{s10,s11} also studied short-range NN
repulsion as stemming from the Goldstone boson (and rho-like)
exchanges on the quark level. It has been shown that these
interactions can substitute traditional OGE mechanism. But whether
OGE or vector meson exchange is the right mechanism for describing
the short range part of the strong interactions, or both of them
are important, is still a challenging problem. To study this
interesting problem, we extend our chiral $SU(3)$ quark model
\cite{s2} to involve vector meson exchanges. As we did before in
the chiral $SU(3)$ quark model, first we fit the masses of baryon
ground states, then the N-N phase shifts are calculated by solving
a Resonating Group Method (RGM) equation to see the effect of
vector meson exchanges. The results show that the $^1S_0$ phase
shifts of one channel calculation are obviously improved in the
extended chiral $SU(3)$ quark model.

The paper is arranged as follows. The theoretical framework of the
extended chiral $SU(3)$ quark model is introduced in section 2.
The N-N phase shifts of $S$, $P$, $D$, and $F$ partial waves
are shown and discussed in section 3. Finally a conclusion is made
in section 4 .

\section{Theoretical framework of the extended chiral $SU(3)$ quark model}
%{(\bf I)}.~
In the extended chiral $SU(3)$ quark model, besides the nonet
pseudo-scalar
 meson fields
and the nonet scalar meson fields, the couplings among vector
meson fields with quarks is also considered. With this
generalization, in the interaction Lagrangian a term of coupling
between quark and vector meson field is included,
\begin{eqnarray}
&{\cal{L}}_{I}^{v}& = -i g_{chv}\overline{\psi}\gamma_{\mu}
\vec{\varphi}_{\mu} \cdot \vec{\tau} \psi -i\frac{f_{chv}}{2
M_p}\overline{\psi}\sigma_{\mu \nu} \partial_{\nu}
\vec{\varphi}_{\mu} \cdot \vec{\tau} \psi.
\end{eqnarray}
Thus the Hamiltonian of the system can be written as
\begin{eqnarray}
& H & =\sum\limits_{i}T_i-T_G+\sum\limits_{i<j}V_{ij},
\end{eqnarray}
and
\begin{eqnarray}
& V_{ij} & =V_{ij}^{conf}+V_{ij}^{OGE}+V_{ij}^{ch},
\end{eqnarray}
where ~$\sum\limits_{i}T_i-T_{G}$ is the kinetic energy of the
system, and $V_{ij}$ includes all the interactions between two
quarks, $V_{ij}^{conf}$ is the confinement potential taken as the
quadratic form,
\begin{eqnarray}
& V_{ij}^{conf} &
=-a_{ij}^{c}(\lambda_{i}^{c}\cdot\lambda_{j}^{c})r_{ij}^2
-a_{ij}^{c0}(\lambda_{i}^{c}\cdot\lambda_{j}^{c}),
\end{eqnarray}
and $V_{ij}^{OGE}$ is the one gluon exchange (OGE) interaction,
\begin{eqnarray}
& V^{OGE}_{ij} & = \frac{1}{4} g_{i}g_{j} (\lambda^{c}_{i} \cdot
       \lambda^{c}_{j})
       \{ \frac{1}{r_{ij}} - \frac{\pi}{2} \delta(\vec{r}_{ij})
       ( \frac{1}{m^{2}_{qi}} + \frac{1}{m^{2}_{qj}} \nonumber \\
   & & + \frac{4}{3} \frac{1}{m_{qi}m_{qj}} (\vec{\sigma_{i}}
        \cdot \vec{\sigma_{j}}) )
        -\frac{1}{4m_{qi}m_{qj}r^{3}_{ij}} S_{ij} \} +
        V^{\vec{\ell}\cdot\vec{s}}_{OGE}
\end{eqnarray}
with
\begin{eqnarray}
& V^{\vec{\ell} \cdot \vec{s}}_{OGE}&  = -\frac{1}{16} g_{i}g_{j}
         (\lambda^{c}_{i} \cdot \lambda^{c}_{j})
         \frac{3}{m_{q_i}m_{q_j}} \frac{1}{r^{3}_{ij}} \vec{L}
         \cdot (\vec{\sigma}_{i}
         + \vec{\sigma}_{j} ) .
\end{eqnarray}
$V_{ij}^{ch}$ represents the interactions from chiral field
couplings. In the extended chiral $SU(3)$ quark model
$V_{ij}^{ch}$ includes scalar meson exchange $V_{ij}^{s}$ ,
pseudo-scalar meson exchange $V_{ij}^{ps}$, and vector meson
exchange $V_{ij}^{v}$ potentials,
\begin{eqnarray}
& V_{ij}^{ch} & = \sum^{8}_{a=0} V_{s_a} (\vec{r}_{ij}) +
\sum^{8}_{a=0}
   V_{ps_a} (\vec{r}_{ij})+ \sum^{8}_{a=0}V_{v_a} (\vec{r}_{ij})~.
\end{eqnarray}
Their expressions are
\begin{eqnarray}
 V_{s_a} (\vec{r}_{ij}) & = & -C(g_{ch}, m_{s_a}, \Lambda_{c})
X_{1}(m_{s_a}, \Lambda_{c}, r_{ij}) (\lambda_{a}(i)\lambda_a (j) ) \nonumber\\
  & + & V^{\vec{\ell} \cdot \vec{s}}_{s_a} (\vec{r}_{ij}),\\
 V_{ps_a}(\vec{r}_{ij}) & = & C(g_{ch}, m_{ps_a}, \Lambda_{c})
\frac{m_{ps_a}^{2}}{12m_{qi}m_{qj}}\{  X_{2}(m_{ps_a},
\Lambda_{c},
r_{ij}) (\vec{\sigma}_{i} \cdot \vec{\sigma}_{j})  \nonumber\\
  & + & \left ( H(m_{ps_a} r_{ij}) - (\frac{\Lambda_{c}}{m_{ps_a}} )^{3}
H(\Lambda_{c} r_{ij} ) \right )\hat{S}_{ij}\}  (\lambda_{a}(i)
\lambda_{a}(j) ) ~~,
\end{eqnarray}
\begin{eqnarray}
 V_{v_a} (\vec{r}_{ij}) & = & C(g_{chv}, m_{v_a}, \Lambda_c)
X_{1}(m_{v_a}, \Lambda_c, r_{ij}) (\lambda_{a}(i)\lambda_a(j)) \nonumber\\
  & + & C(g_{chv}, m_{v_a}, \Lambda_c)
\frac{m_{v_a}^{2}}{6m_{qi}m_{qj}}(1+\frac{f_{chv}}{g_{chv}}
\frac{m_{qi}+m_{qj}}{M_p}+\frac{f_{chv}^2}{g_{chv}^2}\frac{m_{qi}
m_{qj}}{M_p^2}) \nonumber\\
 &\times &\{  X_{2}(m_{v_a}, \Lambda_c,r_{ij})
(\vec{\sigma}_{i} \cdot \vec{\sigma}_{j})  \nonumber\\
  & - &\frac{1}{2} \left ( H(m_{v_a} r_{ij}) - (\frac{\Lambda_c}{m_{v_a}}
)^{3} H(\Lambda_c r_{ij} ) \right )\hat{S}_{ij} \}
(\lambda_{a}(i)\lambda_{a}(j) )
\nonumber\\
  & + & V^{\vec{\ell} \cdot \vec{s}}_{v_a} (\vec{r}_{ij})
\end{eqnarray}
with
\begin{eqnarray}
  V^{\vec{\ell} \cdot \vec{s}}_{s_{a}} (\vec{r}_{ij}) & = & -C(g_{ch},
        m_{s_{a}}, \Lambda_{c})
        \frac{m^{2}_{s_{a}}}{4m_{qi} m_{qj}}
       \{  G (m_{s_{a}} r_{ij} )  \nonumber\\
  & - & (\frac{\Lambda_{c}}{m_{s_{a}}} )^{3}
        G(\Lambda_{c} r_{ij})\}
        (\vec{L} \cdot (\vec{\sigma}_{i} + \vec{\sigma}_{j} ))
        (\lambda_{a}(i)\lambda_{a}(j) ) ,
\end{eqnarray}
and
\begin{eqnarray}
 V^{\vec{\ell} \cdot \vec{s}}_{v_{a}} (\vec{r}_{ij}) & = & -C(g_{chv},
        m_{v_{a}}, \Lambda_{c})
        \frac{3m^{2}_{v_{a}}}{4m_{qi} m_{qj}}(1+\frac{f_{chv}}{g_{chv}}
        \frac{2(m_{qi}+m_{qj})}{3 M_p})\nonumber\\
  &\times&    \{ G (m_{v_{a}} r_{ij} ) - (\frac{\Lambda_{c}}{m_{v_{a}}} )^{3}
        G(\Lambda_{c} r_{ij})\}
        (\vec{L} \cdot (\vec{\sigma}_{i} + \vec{\sigma}_{j} ))
        (\lambda_{a}(i)\lambda_{a}(j) ),
\end{eqnarray}
where
\begin{eqnarray}
C(g, m, \Lambda)& = & \frac{g^{2}}{4\pi}
          \frac{\Lambda^{2} m }{\Lambda^{2} - m^{2} },\\
X_1(m, \Lambda, r)& = & Y(m r) -\frac{\Lambda}{m}Y(\Lambda r),\\
X_2(m, \Lambda, r)& = & Y(m r) -(\frac{\Lambda}{m})^3 Y(\Lambda r),\\
Y(x) & = & \frac{1}{x}e^{-x},\\
H(x) & = & (1+\frac{3}{x}+\frac{3}{x^2}) Y(x),\\
G(x) & = & \frac{1}{x}(1+\frac{1}{x}) Y(x),
\end{eqnarray}

\begin{eqnarray}
\hat{S}_{ij} = 3(\vec{\sigma}_{i} \cdot \hat{r})
        (\vec{\sigma}_{j} \cdot \hat{r})
      -(\vec{\sigma}_{i} \cdot \vec{\sigma}_{j}) ,
\end{eqnarray}
and $M_p$ is a mass scale, taken as proton mass.

In the calculation, $\eta$ and $\eta'$ mesons are mixed by
$\eta_1$ and $\eta_8$, the mixing angle $\theta_{\eta}$ is taken
to be the usual value with $\theta_{\eta}=-23^{o}$. $\omega$ and
$\phi$ mesons consist of
$\sqrt{\frac{1}{2}}(u\overline{u}+d\overline{d})$ and
$(s\overline{s})$ respectively, i.e. they are mixed by $\omega_1$
and $\omega_8$, with the mixing angle $\theta_{\omega}=-54.7^{o}$.

The coupling constant for scalar  and pseudo-scalar chiral field
coupling, $g_{ch}$, is determined according to the relation
\begin{eqnarray}
\frac{g_{ch}^2}{4\pi} = \frac{9}{25}~ \frac{m_u^2}{M_N^2}~
\frac{g_{NN\pi}^2}{4\pi}~,
\end{eqnarray}
where ${g_{NN\pi}^2}/{4\pi}=13.67$~ is taken from the experimental
value. $g_{chv}$ and  $f_{chv}$ are the coupling constants for
vector coupling and tensor coupling of the vector meson field
respectively. In the study of nucleon resonance transition
coupling to vector meson, Riska et al
\cite{s13}~took~$g_{chv}=3.0$ and neglected the tensor coupling
part. From the one boson exchange theory on baryon level, we can
also obtain these two values according to the $SU(3)$ relation
between quark and baryon levels. For example,
\begin{eqnarray}
g_{chv} = g_{NN\rho}~,\\
f_{chv} = \frac{3}{5}(f_{NN\rho}-4 g_{NN\rho})~.
\end{eqnarray}
In the Nijmegen model D, $g_{NN\rho}=2.09 $ and $f_{NN\rho}=17.12
$. From Eqs. (20) and (21), we get $g_{chv} = 2.09 $ and $f_{chv}
= 5.26 $. Such information is useful for adjusting the coupling
constants of vector meson exchanges on quark level. All meson
masses, $m_{ps}$, $m_{s}$, and $m_{v}$, are taken to be the
experimental values, except the mass of $\sigma$ meson,
$m_{\sigma}$. According to the dynamical vacuum spontaneous
breaking mechanism, its value should satisfy \cite{s14}
\begin{eqnarray}
m_{\sigma}^2= (2 m_u)^2 + m_{\pi}^2.
\end{eqnarray}
This means that it can be regarded as almost reasonable when the
value of $m_{\sigma}$ is located in the range of around $550\sim
650$ MeV. In our calculation, we treat it as an adjustable
parameter by fitting the binding energy of deuteron and in most
other cases within this reasonable range. The cut-off mass,
$\Lambda$, indicating the chiral symmetry breaking scale
\cite{s15,s16} is taken to be $\Lambda^{-1}=0.18fm$. Once the
parameters of chiral field are fixed, the coupling constant of OGE
$g_u (\alpha_s=g_u^2)$, $g_s$, and the strength of confinement
potential $a_{uu}, a_{us}, a_{ss}$ can be determined by fitting
the baryon masses and by their stability conditions. All the
parameters used are listed in Table 1. Here we should mention that
there are only one coupling constant for vector coupling of vector
meson $g_{chv}$  and one for tensor coupling $f_{chv}$  in our
model. The number of adjustable parameters are largely reduced in
comparison to the one boson exchange theory on baryon level, in
which different coupling constants were adopted for $\rho, K^*$,
$\omega$, and $\phi$ exchanges.

Equipped with the extended chiral $SU(3)$ quark model with all the
parameters determined, two baryon systems on dynamical quark level
will be studied by solving the RGM equation with the Hamiltonian
given by Eqs.(2)-(19).

In the RGM calculation, the trial wave function is taken to be
\begin{equation}
\Psi_{ST}=\sum\limits_{i}c_i\Psi_{ST}^{(i)}(\vec{s_i})
\end{equation}
with
\begin{equation}
\Psi_{ST}^{(i)}(\vec{s_i})={\cal{A}}(\phi_{A}(\vec{\xi}_{1},\vec{\xi}_{2})
_{S_{A}T_{A}}\phi_{B}(\vec{\xi}_{4},\vec{\xi}_{5})_{S_{B}T_{B}}
\chi(\vec{R}_{AB}-\vec{s}_{i}) {\cal{R}}_{CM}(\vec{R}_{CM}))_{ST},
\end {equation}
where A and B describe two clusters, and $\phi$, $\chi$, and
$\cal{R}$ represent internal, relative, and center of mass motion
wave functions, respectively, $\vec{s_i}$ is the generator
coordinate, and $\cal{A}$ is the anti-symmetrization operator
\begin {equation}
{\cal{A}}=1-\sum\limits_{i\in A,j\in B}P_{ij},
\end{equation}
where $P_{ij}$ is the permutation operator of the $i$-th and
$j$-th quarks. The partial wave phase shifts of $N$-$N$ scattering
are studied, and the results will be shown in the next section.

\section{Results of N-N phase shifts and discussions}
%{(\bf II)}.~
We calculated the N-N phase shifts of different partial
$S$,$P$,$D$,$F$, and $G$ waves using the extended chiral $SU(3)$
quark model. The results of the phase shifts are shown in Figs.1 -
4. The corresponding deuteron binding energies $B_{deu}$ are
listed in Table 1. For comparison with results from different
models, the results of chiral $SU(3)$ quark model without vector
meson exchanges \cite{s2} are drawn with short-dashed curves,
while long dashed and solid curves are those obtained from the
extended chiral $SU(3)$ quark model with two different sets of
parameters. In the case of set I, the tensor coupling of the
vector meson exchanges is not considered, namely
$f_{chv}/g_{chv}=0$, while the tensor coupling of vector meson
exchanges is included in the case of set II, where
$f_{chv}/g_{chv}=2/3$.  As mentioned before, we adjust the mass of
$\sigma$ meson $m_{\sigma}$ to fit the binding energy of deuteron.
From Table 1, one can see that $m_{\sigma}$ is almost located in
the reasonable range $550\sim 650$ MeV for all these three cases
when the binding energy of deuteron $B_{deu}$ is fitted.
\begin{table}
\caption{Model parameters and the corresponding binding energies
of deuteron. The meson masses and
%parameters
the cut-off mass $\Lambda$ are $m_{\pi}=138 MeV, m_{K}=495 MeV,
m_{\eta}=548 MeV, m_{\eta'}=958 MeV, m_{\sigma'}= m_{\kappa}=
m_{\epsilon}=980 MeV, m_{\rho}=770 MeV, m_{K^*}=892 MeV,
m_{\omega}=782 MeV, m_{\phi}=1020 MeV $ and $\Lambda=1100 MeV$.}
\begin{small}
\begin {center}
\begin{tabular}{|c|c|c|c|}
\hline
%($~$)
                       & chiral $SU(3)$ quark model &
\multicolumn{2}{|c|}{extended  chiral $SU(3)$ quark model}   \\
\hline
                       &          & ~~~~~~ set I ~~~~~&  set
                        II
              \\ \hline
$b_u (fm)$             & 0.5      & ~~~~~~ 0.45  ~~~~   & 0.45      \\
$g_{nn\pi}$            & 13.67    & 13.67    & 13.67     \\
$g_{ch}$               & 2.621    & 2.621    & 2.621     \\
$g_{chv}$              & 0        & 2.351    & 1.972     \\
$f_{chv}/g_{chv}$      & 0        & 0        & 2/3       \\
$m_{\sigma}(MeV)$      & 595      & 535      & 547       \\
$g_u$                  & 0.886    & 0.293    & 0.399     \\
$\alpha_s (g_u^2)$     & 0.785    & 0.086    & 0.159     \\
$a_{uu}(MeV/fm^2)$     & 48.1     & 48.0     & 42.9      \\
\hline
                       &          &          &           \\
$B_{deu}(MeV)$         & 2.13     & 2.19     & 2.14      \\
\hline
\end{tabular}
\end{center}
%\end{flushleft}
\end{small}

\end{table}

From Table 1 and Figs.1-4, some interesting results are shown: (1)
The one channel calculation of $^1S_0$ phase shifts is improved in
the extended chiral $SU(3)$ quark model in comparison to that
obtained from chiral $SU(3)$ quark model, especially in the case
of set II, in which the tensor coupling of the vector meson
exchange is considered though $f_{chv}/g_{chv}=2/3$ is not so
large as that in the Nijmegen model D. Originally in the chiral
$SU(3)$ quark model, we need to consider $(NN)_{LSJ=000}$ and
$(N\Delta)_{LSJ=220}$ channel coupling to get reasonable $^1S_0$
phase shifts \cite{s2}. In that case, since the non-diagonal
matrix elements between these two channels offered by pion tensor
interactions are quite large, the coupling effect becomes very
important. In this work, we also performed coupled channel
calculation for $^1S_0$ state. The results are shown in Figs.
5a-5c, in which dashed curve represents the one channel
calculation results, and solid curves are those with
$(NN)_{LSJ=000}$ and $(N\Delta)_{LSJ=220}$ channel coupling.
Fig.5a is the result of the chiral $SU(3)$ quark model. It shows
that the coupled channel effect is quite important in the chiral
$SU(3)$ quark model\cite{s2}. Figs. 5b and 5c are results of the
extended  chiral $SU(3)$ quark model with parameters of set I and
set II. In the extended chiral $SU(3)$ quark model, since the
tensor parts of vector mesons ($\rho$ and $\omega$) have opposite
sign to that of pion in the $T=1$ case, the non-diagonal matrix
elements between $(NN)_{LSJ=000}$ and $(N\Delta)_{LSJ=220}$
channels can be reduced due to the vector meson contribution, so
that the coupled channel effect becomes smaller. But for all these
three cases, the $^1S_0$ phase shifts of the coupled channel
calculation are a little bit higher than the corresponding
experimental values. (2)~ The $^3S_1$ phase shifts of different
models are almost the same, and all of them are in good agreement
with experimental data. To get the right trend of the phase shift
versus scattering energy, the size parameter $b_u$ is taken with
different values for these two models. $b_u = 0.50 fm$ in the
chiral $SU(3)$ quark model, and $b_u = 0.45 fm$ in the extended
chiral $SU(3)$ quark model. This means that the bare radius of
baryon becomes smaller when more meson clouds are included in the
model. This physical picture looks reasonable. (3)~When the vector
meson field coupling is considered, the coupling constant of OGE
is greatly reduced by fitting the mass difference between $\Delta$
and $N$. For both set I and II, $g_{u}^2 (\alpha_s)< 0.2$, which
is much smaller than the value $(0.785)$ of chiral $SU(3)$ quark
model (see Table 1). It means that  the OGE interaction is rather
weak in the extended chiral $SU(3)$ quark model. Instead, the
vector meson exchanges play an important role for the short range
interaction between two quarks. Hence, mechanisms of the
quark-quark short range interactions of these two models are
totally different. In the chiral $SU(3)$ quark model, the short
range interaction is dominantly from OGE, and in the extended
chiral $SU(3)$ quark model, it is from combined effect of ps- and
vector meson exchanges. ~(4)~To see the contributions from
different meson exchanges in the N-N interactions in our extended
chiral $SU(3)$ quark model, we print out the diagonal generator
coordinating method (GCM) matrix elements for $\pi$, $\rho$ and
$\omega$ mesons respectively. As an example, Fig. 6 gives the
potentials of $\pi$, $\rho$ and $\omega$ mesons  for the $^1S_0$
state (for set I). The $\pi$, $\rho$ and $\omega$ meson exchange
potentials are drawn with solid, long-dashed and short-dashed
curves, respectively. One can see that the $\omega$ meson exchange
offers repulsion not only in the short range region, but also in
medium range part. This property is different from that of $\pi$
meson, which only contributes repulsive core. ~(5)~As for the
coupling constants of the vector meson exchange $g_{chv}$ and
$f_{chv}$, when we take $f_{chv}/g_{chv}=0$, we get
$g_{chv}=2.35$, which is a little bit smaller than the value used
by Riska et al \cite{s13}, but slightly larger than the value
obtained from the $NN \rho$ coupling constant of Nijmegen model D
\cite{s9}. When $f_{chv}/g_{chv}=2/3$, we get $g_{chv}=1.97$ (see
Table 1), which is very closed to the value obtained from the
Nijmegen model D, but $f_{chv}/g_{chv}$ are much smaller than that
from Nijmegen model D. The results indicate that the coupling
constant of vector meson exchange on quark level is much weaker
than the corresponding coupling constant on baryon level. Since on
the quark level the size effect and the quark exchanges between
two nucleon clusters also contribute short range repulsion,  the
smaller coupling constant of vector meson exchange on quark level
can easily be understood.
%%%%%%%%%%%%%%%
From the phase shift calculations, one can also see that the
results are not very sensitive to $f_{chv}/g_{chv}$. For both case
I and II, the calculated phase shifts are almost consistent with
the experimental data. When $f_{chv}/g_{chv}=2/3$, it is slightly
helpful to increase the one channel $^1S_0$ phase shifts. It seems
that on quark level the tensor coupling part of the vector meson
exchanges is not very important in explaining the $S$ wave phase
shifts. But for the $^3P_0$,$^3P_2$, $^3D_2$, and $^3D_3$ partial
waves, the calculated results are still not good enough, because
in the extended chiral $SU(3)$ quark model, there is only one
coupling constant $g_{chv}$ for the vector coupling of the vector
meson exchanges, and one coupling constant $f_{chv}$ for the
tensor coupling part. In the one boson exchange model on
baryon level,
such as Nijmegen model, and the work on quark level
\cite{s12}, however, the coupling constants of $\rho$ and $\omega$
mesons are taken to be quite different. Especially, the tensor
coupling constant $\omega$, $f_{NN\omega}$ are quite large
\cite{s9}. If we take different $g_{chv}$ and $f_{chv}$ for $\rho$
and $\omega$ exchanges, possibly the results of $^3P_0$,$^3P_2$,
$^3D_2$, and $^3D_3$ partial waves can be improved.

\section{ Conclusions}

The vector meson exchange effect in N-N scattering processes on
quark level is studied in the extended chiral $SU(3)$ quark model.
The differences between the chiral $SU(3)$ quark model and the
extended chiral $SU(3)$ quark model are discussed in detail. It is
found that in the extended chiral $SU(3)$ quark model, the phase
shifts of $^1S_0$ and $^3S_1$ waves can be fitted rather well. All
other partial waves  calculated are also consistent with the
experimental results. In the extended chiral $SU(3)$ quark model,
the strength of OGE interaction is  greatly reduced and the short
range NN repulsion is due to combined effect of ps- and vector
meson exchanges, which also results in smaller size parameter
$b_u$. The tensor coupling constant of the vector meson exchange
$f_{chv}$ also becomes smaller than that on baryon level. All of
these features shown above are reasonable and helpful in better
understanding the short range mechanism of the quark-quark
interactions.

\vspace{1.0cm} \noindent {\bf Acknowledgement}

We are in debt to Prof. Shen Pengnian for stimulating discussions.

\newpage
\begin{figure}[h!]
\epsfig{figure=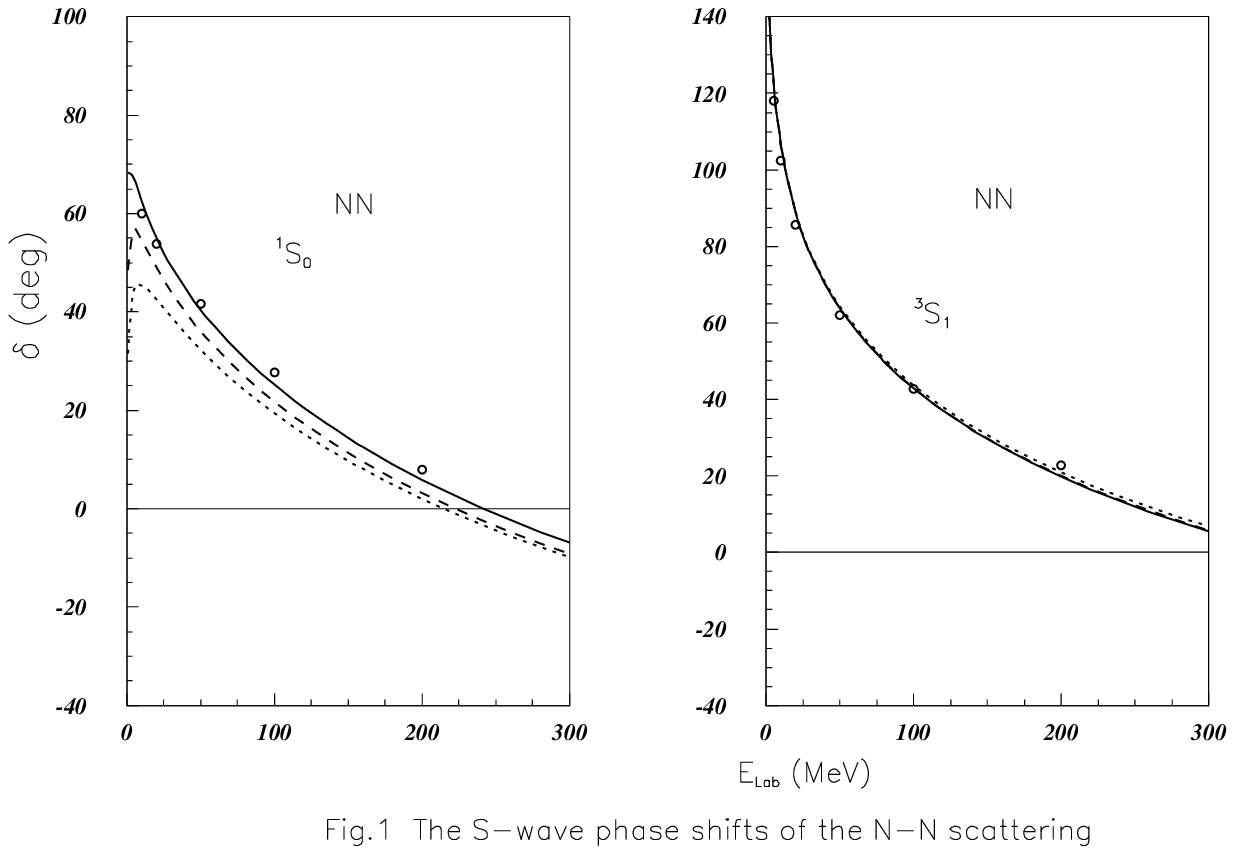}
\end{figure}

\begin{figure}[h!]
\epsfig{figure=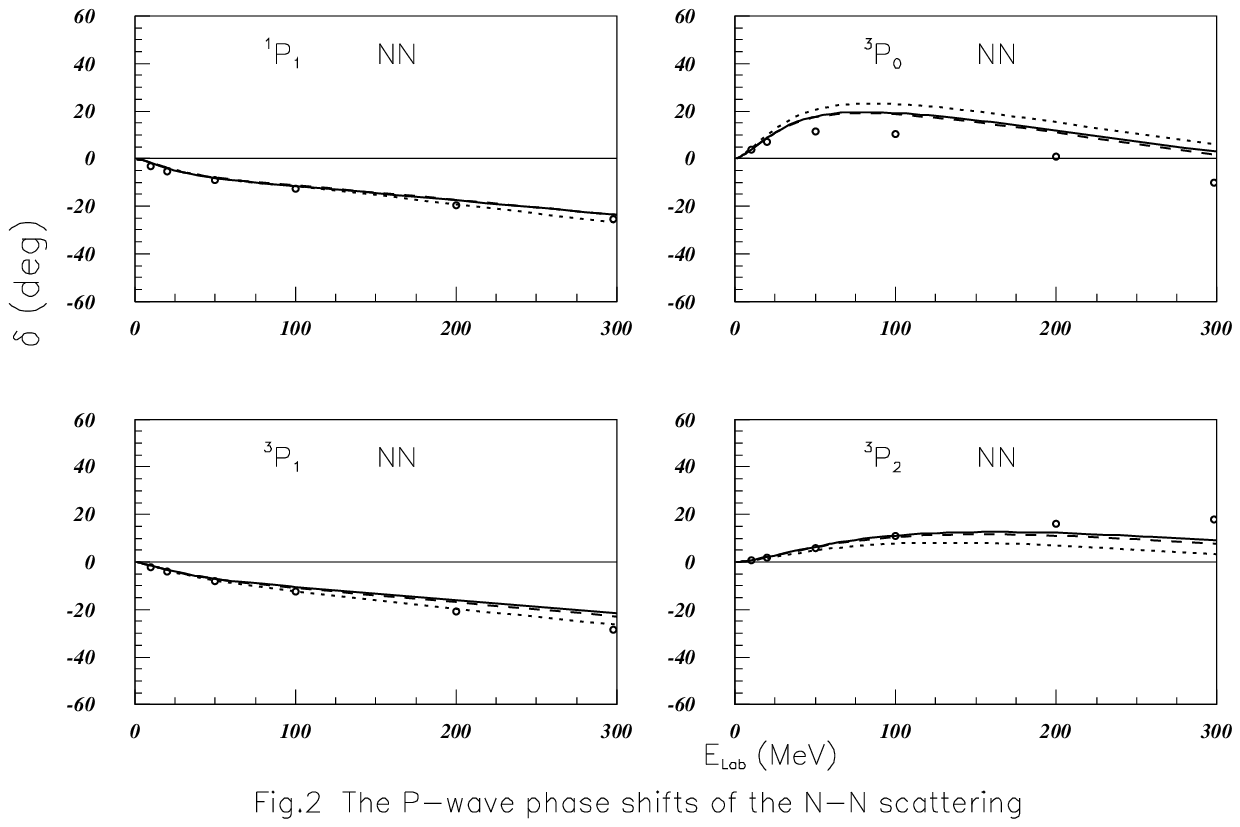}
\end{figure}

\begin{figure}[h!]
\epsfig{figure=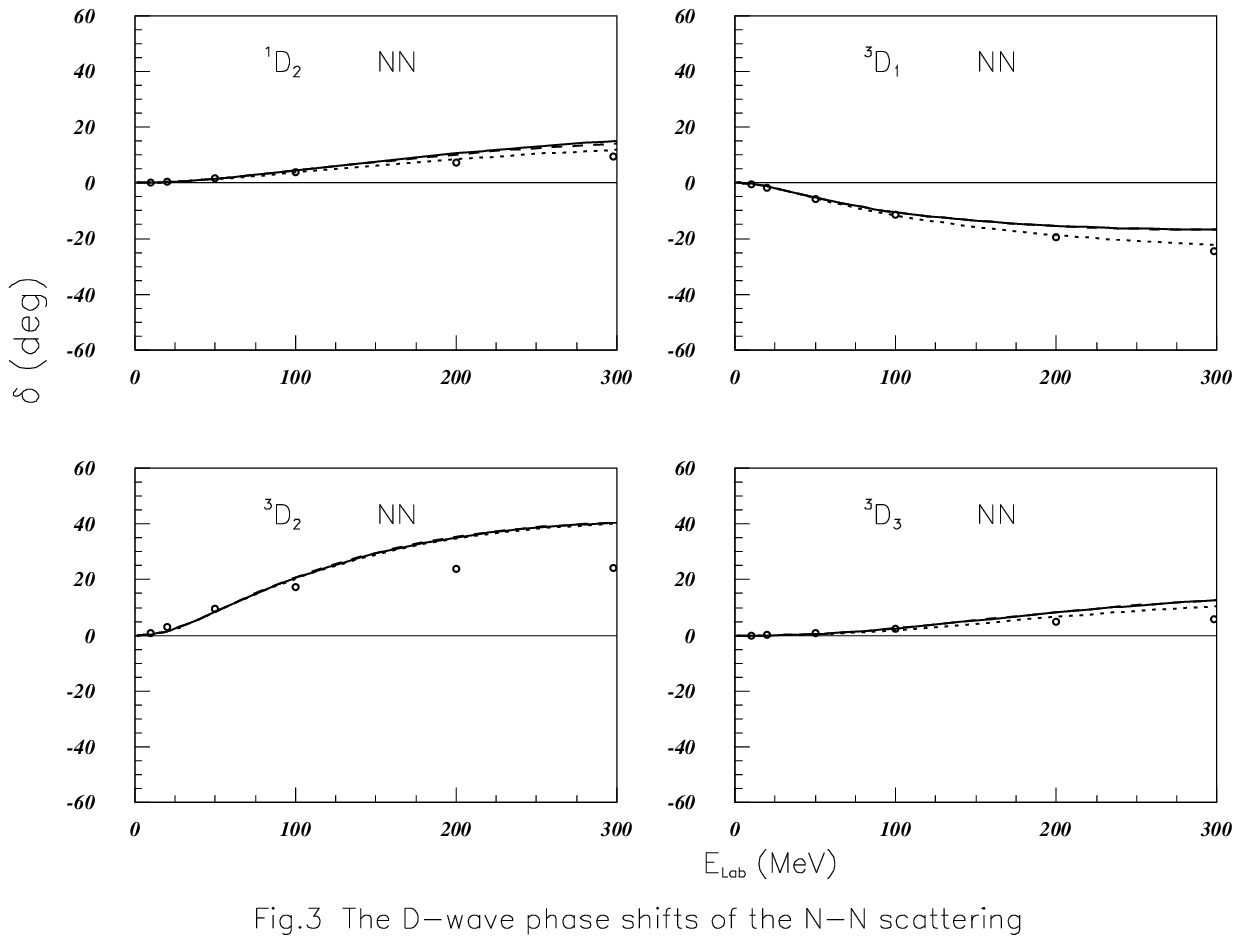}
\end{figure}

\begin{figure}[h!]
\epsfig{figure=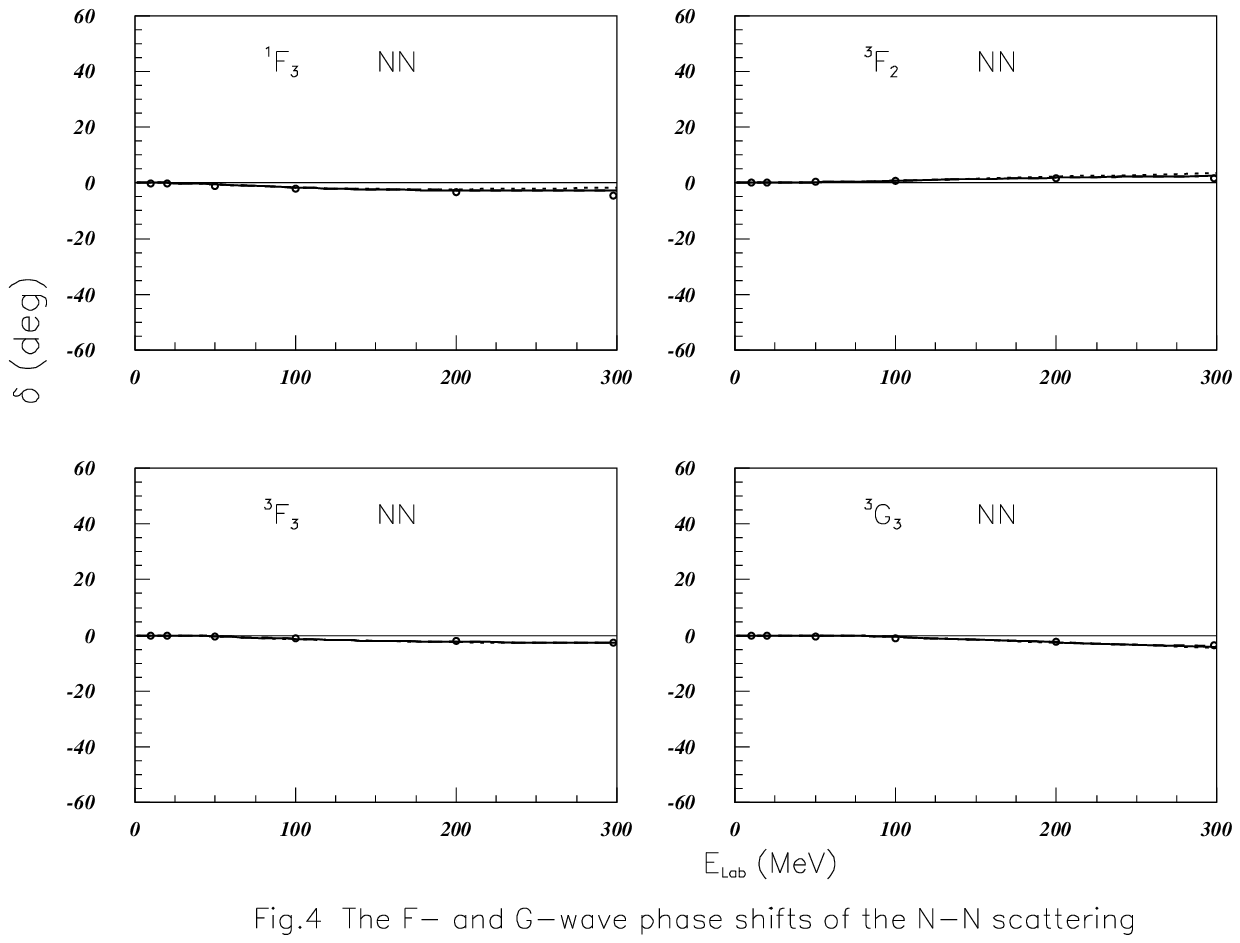}
\end{figure}

\begin{figure}[h!]
\epsfig{figure=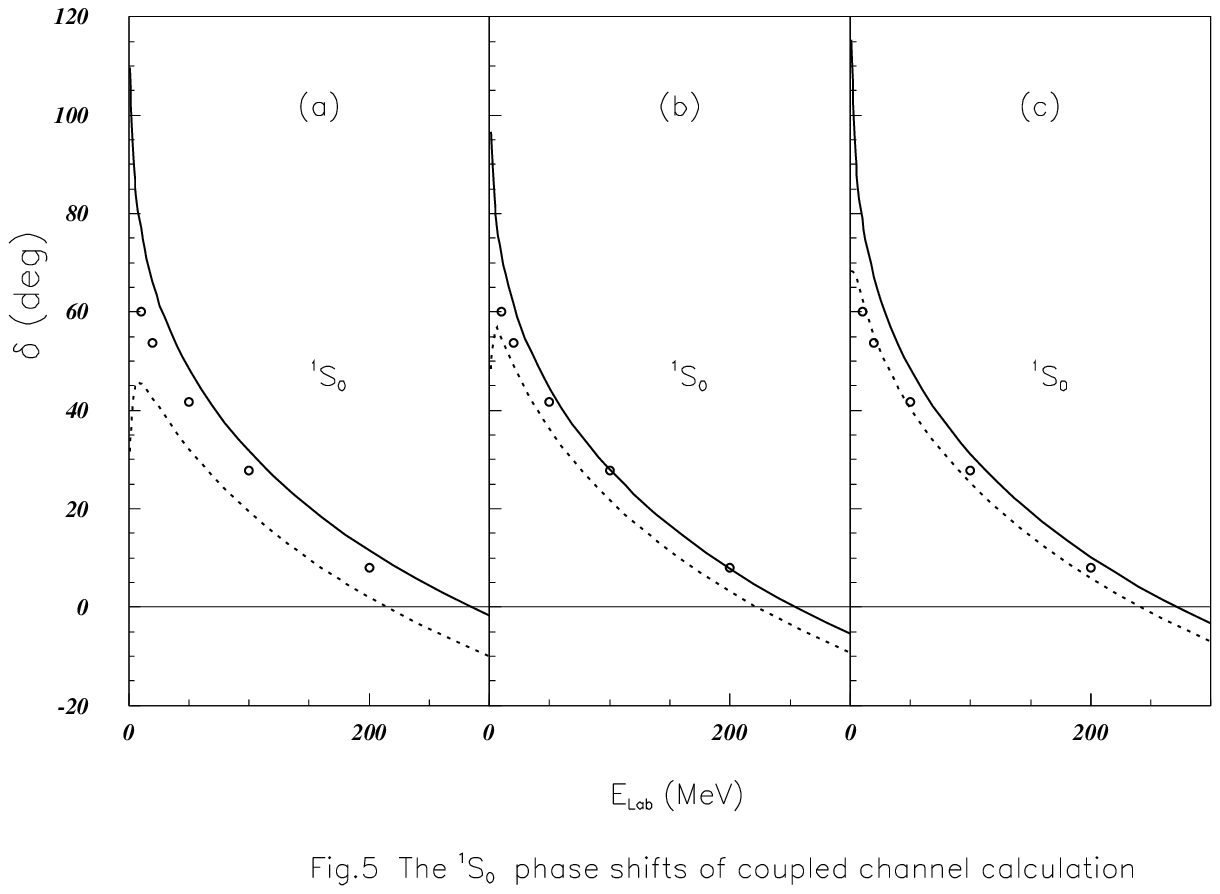}
\end{figure}

\begin{figure}[h!]
\epsfig{figure=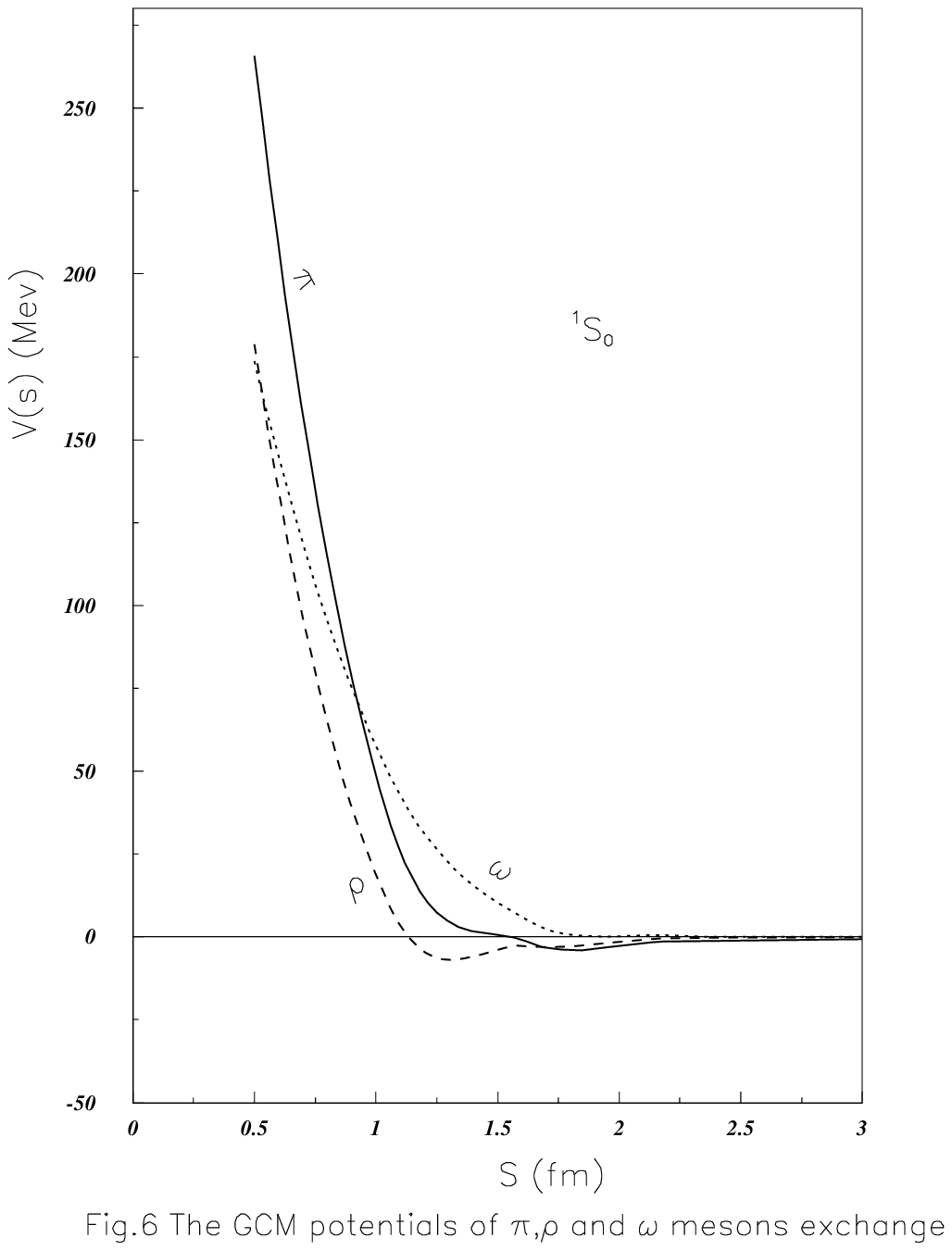}
\end{figure}

\end{document}